\documentclass[12pt]{iopart}
\usepackage[utf8]{inputenc}
\usepackage{iopams}
\usepackage{dsfont}
\usepackage{float}
\usepackage{cancel}
\usepackage{subfiles}
\usepackage[caption=false]{subfig}
\usepackage{nicefrac}
\usepackage{tabularx}
\usepackage{braket}
\usepackage{graphicx}
\usepackage{color}
\usepackage[usenames,dvipsnames,svgnames,table]{xcolor}
\usepackage{bm}
\usepackage{ulem}
\usepackage{pdfpages}

\newcommand{\binom}[2]{{#1\choose #2}}
\newcommand{\hc}{\mathrm{h.c.}}
\newcommand{\comm}[2]{\left[#1, #2\right]}

\newcommand{\lrangle}[1]{\left\langle #1 \right\rangle}
\newcommand{\op}[1]{\hat{#1}}
\newcommand{\supop}[1]{\mathbb{#1}}
\newcommand{\lvl}{\supop{L}}
\newcommand{\spin}[2]{\op{\sigma}_{#2}^{\mathrm{#1}}}

\newcommand{\nnref}[3]{\hyperref[#1]{#2\ref*{#1}#3}}

\newcommand{\stkout}[1]{\ifmmode\text{\sout{\ensuremath{#1}}}\else\sout{#1}\fi}

\newcommand{\nndel}[1]{{\color{red} \stkout{#1}}}
\newcommand{\nncomment}[1]{{\color{blue} \textit{(#1)}}}

\renewcommand{\nndel}[1]{}
\renewcommand{\nncomment}[1]{}

\usepackage{hyperref}

\begin{document}
\submitto{\NJP}

\title[Long-time memory in a localizable central spin problem]{
Long-time memory effects in a localizable central spin problem
}
\author{Nathan Ng}
\address{Department of Physics, University of California, Berkeley, CA 94720, USA}
\author{Eran Rabani}
\address{Department of Chemistry, University of California, Berkeley, CA 94720, USA}
\address{Materials Sciences Division, Lawrence Berkeley National Laboratory, Berkeley, CA 94720, USA}
\address{The Sackler Center for Computational Molecular and Materials Science, Tel Aviv University, Tel Aviv 69978, Israel}
\date{\today}

\begin{abstract}
We study the properties of the Nakajima-Zwanzig memory kernel for a qubit immersed in a many-body localized (i.e.\ disordered and interacting) bath.
We argue that the memory kernel decays as a power law in both the localized and ergodic regimes, and show how this can be leveraged to extract $t\to\infty$ populations for the qubit from finite time ($J t \leq 10^2$) data in the thermalizing phase.
This allows us to quantify how the long-time values of the populations approach the expected thermalized state as the bath approaches the thermodynamic limit.
This approach should provide a good complement to state-of-the-art numerical methods, for which the long-time dynamics with large baths are impossible to simulate in this phase.
Additionally, our numerics on finite baths reveal the possibility for unbounded exponential growth in the memory kernel, a phenomenon rooted in the appearance of exceptional points in the projected Liouvillian governing the reduced dynamics.
In small systems amenable to exact numerics, we find that these pathologies may have some correlation with delocalization.
\end{abstract}

\maketitle

\section{Introduction}
Central spin models are ubiquitous in physical and chemical settings, from electrons with hyperfine coupling to nuclear spins inside quantum dots~\cite{Schliemann2003,Witzel2010,Chekhovich2013,Cogan2018}, to nitrogen-vacancy centers in diamond~\cite{Hanson2008,deLange2012,Bauch2020}.
Depending on the couplings in these systems, the central spin may have long-lived, slow decaying dynamics suitable for quantum information applications.
The role of the bath in these cases is relegated to modeling decoherence, and has not traditionally been considered to be important.
The bath is usually taken to be non-interacting, an assumption which has proven fruitful in the development of analytical~\cite{Dukelsky2004,Claeys2018,Villazon2020, Bortz2007,Bortz2010,Barnes2012,Hall2014} and numerical~\cite{Wolf2014, Wang2012, Diosi1997, Orth2013, Lindoy2018, Faribault2013} techniques.
As such, these classes of baths--whether composed of bosons~\cite{Leggett1987} or spins~\cite{Prokofev2000}--are by now reasonably well understood~\cite{Prokofev2000,Khaetskii2003,Coish2004,Chen2007,Barnes2011,Erbe2012,Zhou2020,Nepomechie2018,Jing2018}.

Recent research has brought new focus to modifications of the bath by adding, for example, intra-bath interactions and disorder.
With these additions, the bath alone can exhibit novel dynamical phases such as many-body localization (MBL), which serves as a basis for nonergodicity in generic systems with strong disorder.
Upon coupling to a bath, the long-ranged mediated interactions between constituents of the bath can push the bath towards delocalization. 
Recent work~\cite{Hetterich2018, Ponte2017} has shown that a single qubit coupled centrally to a 1D MBL spin chain can preserve localization provided that the magnitude of the central coupling decays fast enough with the size of the bath.
Delocalization can be achieved by a sufficiently strong magnitude of central coupling, which the authors of \cite{Hetterich2018} took to be signaled by quantum chaotic energy level statistics.
However, it was noted in \cite{Ponte2017} that the nature of the delocalized phase is unclear, as it could be nonergodic.
This could be reflected in the long-time value of the central qubit's population not reaching the thermal expectation but, as was found in \cite{Villazon2020} studying integrable central spin models perturbed away from integrability, limitations of bath size prevent a definitive conclusion.
An impediment is that at strong couplings, analytical and numerical approaches become scant due to the presence of interactions in the bath and to the star-like geometry of problem.

The added complexity has a drawback in that such systems quickly become intractable computationally, even for small bath sizes of $\sim O(30)$ degrees of freedom.
This is due to the exponential increase of states in the Hilbert space that are involved in the dynamics.
Moreover, large intra-bath interactions and disorder can radically change the timescales of the bath and invalidate perturbative approaches to bath dynamics.
In this context, the case of MBL (in one dimension) is special in that it allows for a non-perturbative description in terms of 'l-bits'~\cite{Huse2014, Nandkishore2015, Imbrie2017, Abanin2019}.
Owing to this and to the slow growth of entanglement entropy~\cite{Znidaric2008, Bardarson2012}, dynamics in the localized phase of MBL systems are by now well explored numerically and analytically~\cite{BarLev2015, DeTomasi2019, Alet2018, Abanin2019, Chanda2020}; however, these approaches generally fail on the ergodic side of the transition.

The dynamics of extended, thermalizing many-body systems are typically very difficult to simulate exactly due to the rapid growth of entanglement.
This is, for example, the limiting factor in methods based on a tensor network ansatz for the wavefunction in which the bond dimension bounds the amount of entanglement entropy that can be captured.
A reasonable strategy then would be to extend the timescale of the converged simulation using information that can be computed on the timescales before the breakdown of the numerical method.
Such an approach had been used successfully in the past to find the steady state behavior of quantum impurity systems~\cite{Cohen2011, Cohen2013, Wilner2014}, and to show the existence of bistability in the Anderson-Holstein model~\cite{Wilner2013}.
In those applications, the nontrivial dynamics of the impurity could be described exactly using a memory kernel, derived using the projection operator formalism described by Nakajima, Zwanzig and Mori~\cite{Nakajima1958, Zwanzig1960, Mori1965}.

While the Nakajima-Zwanzig theory is formally exact, it is oftentimes more demanding than other formalisms to describe the dynamics because of the time-nonlocal memory kernel that naturally arises in their approach.
It only becomes computationally useful if the nonlocality can be restricted, e.g.\ large timescale separation between bath and system dynamics lending to Markovian approximations, or if memory kernel decays sufficiently rapidly such that it can be truncated for times $\geq t_c$, where $t_c$ is the cutoff time.

The use of memory kernels to study dynamics in central spin systems has seen various degrees of success~\cite{Coish2004, Coish2010, Barnes2011, Barnes2012}.
For analytical tractability, such studies are usually restricted to noninteracting baths without disorder and the memory kernel is expanded perturbatively.
In the cases where such expansions are valid, it has been found that the memory exhibits nonexponential decay at long times, with long-time averaged population consistent with a nonergodic dynamics~\cite{Coish2004}.
However within the perturbative approach it is found that at higher orders of the expansion, the memory kernel can display secular (unbounded) growth~\cite{Coish2004, Barnes2011}.
In this work, we shall go beyond these approaches, taking into account the presence of bath-bath interactions along with random disorder and directly computing the memory kernel, therefore bypassing the possibility of pathological behaviors in the perturbation.

In this paper we study the memory kernel of a two-level system immersed in a bath modeled by a many-body localizable spin chain.
We do so with two goals in mind: to assess the feasibility of extending the system dynamics from short time calculations when analytical and direct numerical approaches to compute the system dynamics fail (i.e.\ on the thermalizing side of the MBL transition); and to understand how interactions and disorder in the bath affect the memory kernel in properties such as timescales and tail behavior.

To this end, we will work with a previously studied model~\cite{Hetterich2018} of a qubit ($\op{\tau}^{x,y,z}$) coupled to a disordered Heisenberg chain of $L$ spins-$1/2$ ($\spin{x,y,z}{i}$):
\begin{eqnarray}
 \label{eq:model}
  \op{H} &= \op{H}_S + \op{H}_B + \op{V}, \\
  \op{H}_S &= \Omega \op{\tau}^z \nonumber\\
  \op{H}_B &=  \sum_{i=1}^L \frac{h_i}{2} \spin{z}{i} + J \sum_{i=1}^L \frac{1}{4} \spin{z}{i}\spin{z}{i+1} + \frac{1}{2} \left( \spin{+}{i}\spin{-}{i+1} + \hc \right) \nonumber \\
  \op{V} &= \frac{\gamma}{L} \sum_{i=1}^L \frac{1}{4} \spin{z}{i} \op{\tau}^z + \frac{1}{2} \left( \spin{+}{i}\op{\tau}^- + \hc \right) \nonumber
\end{eqnarray}
where the $\op{\tau}$ and $\op{\sigma}$ are Pauli matrices. 
The bath Hamiltonian $\op{H}_B$ corresponds to the disordered, isotropic Heisenberg chain, where we take $J=1$.
The system-bath coupling terms $\op{V}$ are likewise given by the Heisenberg interaction, with magnitude scaling as $\gamma/L$ to ensure that localization can occur for finite $\gamma$.
We shall refer to $\gamma$ as the strength of the central coupling.
The random longitudinal fields $h_i$ are drawn independently and uniformly from $[-W,W]$.
The data we present here will be restricted to $W/J = 6$, chosen such that the bath is localized for $\gamma = 0$ and experiences a central coupling-induced delocalization~\cite{Hetterich2018} around $\gamma \approx 5$.
Finally, the magnetic field is set to $\Omega = 0$.
Thus the qubit has no intrinsic dynamics and is instead entirely dependent on the magnitude of the Overhauser field it experiences from the bath. 

As noted in \cite{Hetterich2018}, the interacting central spin problem of \nnref{eq:model}{Eq.~(}{)} can be realized in dipolar spin ensembles.
Other platforms that allow for experimental realizations of this model include programmable quantum simulators~\cite{Zhukov2018}, NMR experiments with triphenylphosphine~\cite{Niknam2020}, or in superconducting qubit circuits~\cite{Song2017}; these approaches offer a high degree of control, for example by enabling the control of intrabath interactions, random disordered Zeeman fields, and the strength of coupling $\gamma$.

The structure of this paper is as follows: we shall first define the memory kernel for reduced dynamics and consider the role of disorder averaging; then we shall analyze the physics underlying the memory kernel at short, intermediate, and long times; and finally we shall discuss the potential for the memory to be used to augment short-time experimental or numerical data. 

\subsection{The Nakajima-Zwanzig equation}
\label{sec:nakajima_zwanzig}
We quickly review the basics of the projection operator approach to generalized quantum master equations.
Any given Hamiltonian can be split into contributions $\op{H}_S$ acting only on the system, $\op{H}_B$ acting only on the bath, and $\op{V}$ coupling the two.
We will use the term ``bath'' as a shorthand for the set of physical degrees of freedom surrounding the central qubit. In particular, we do not assert the character of the bath to be unchanged by coupling to the system.
To each of the three aforementioned operators is associated a corresponding Liouvillian superoperator ($\lvl_S \cdot \, \equiv [\op{H}_S, \cdot]$, $\lvl_B \cdot \, \equiv [\op{H}_B, \cdot]$, $\lvl_V \cdot \, \equiv [\op{V}, \cdot]$) generating dynamics for the density matrix
\begin{eqnarray}
  i\frac{d\op{\rho}}{dt} &= i\frac{d}{dt} e^{-i\lvl t}\op{\rho}_0 = \lvl \op{\rho}(t) \equiv (\lvl_S + \lvl_B + \lvl_V) \op{\rho}(t).
\label{eq:liouville}
\end{eqnarray}
Oftentimes one is interested only in the dynamics of the system, in which case the bath degrees of freedom can be projected out by tracing over the bath on both sides of the equation, where the bath trace is
\begin{eqnarray}
  \Tr_B \op{O} &= \sum_{s, s'}^{\mathrm{dim} \mathcal{H}_S} \sum_b^{\mathrm{dim} \mathcal{H}_B} |s\rangle \langle s'| \, \langle s\otimes b| \op{O} | s'\otimes b\rangle.
\label{eq:partialTrace}
\end{eqnarray}
This is used to define the system reduced density matrix,
\begin{eqnarray}
\op{\rho}_S(t) = \Tr_B \op{\rho}(t).
\end{eqnarray}
We shall additionally assume that the initial state is factorized, i.e.\ $\op{\rho}_0 = \op{\rho}_{S,0} \otimes \op{\rho}_B$.
By taking the bath trace defined in \nnref{eq:partialTrace}{(}{)} on both sides of \nnref{eq:liouville}{(}{)} and using $\Tr_B \lvl_B = 0$, we arrive at the exact expression
\begin{eqnarray}
  i\frac{d}{dt}\op{\rho}_{S}(t)  &= \lvl_S \op{\rho}_{S}(t) + \Tr_B \left( \lvl_V  e^{-i \lvl t} (\op{\rho}_{S,0}\otimes\op{\rho}_B) \right),
\end{eqnarray}
which is an equation of motion for $\op{\rho}_{S}(t)$ that explicitly depends on knowledge of the time evolution of the full system and bath.
This equation of motion can be closed, i.e.\ involving only $\op{\rho}_{S}(t)$, by using Dyson's identity (see \cite{Zwanzig1961, Singh2016}):
\begin{eqnarray}
  i\frac{d}{dt}\op{\rho}_{S}(t)  &= \lvl_S \op{\rho}_{S}(t) - i \int_0^t d\tau \supop{K}(t - \tau) \op{\rho}_S(\tau).
\end{eqnarray}
The memory kernel superoperator is formally defined as 
\begin{eqnarray}
\supop{K}(t) \op{\rho}_S = \Tr_B \left( \supop{P} \lvl \supop{Q} e^{-i\supop{Q}\lvl\supop{Q}t} \supop{Q} \lvl \op{\rho}_S \otimes \op{\rho}_B\right).
  \label{eq:laplace_memory}
\end{eqnarray}
In the above equation, the projection superoperator is taken to be $\supop{P} \cdot \, \equiv \Tr_B(\,\cdot\,) \otimes \rho_B$ and $\supop{Q} = \supop{I} - \supop{P}$ is its complement.
It is useful to define the system reduced propagator (superoperator) such that
\begin{eqnarray}
\supop{U}_S(t) \op{\rho}_{S,0} \equiv \op{\rho}_S(t) = \Tr_B (e^{-i \lvl t} \op{\rho}_{S,0} \otimes \op{\rho}_B).
\end{eqnarray}
Knowledge of $\supop{U}_S$ allows for the generation of $\op{\rho}_S(t)$, and lets us write a Nakajima-Zwanzig equation~\cite{Kidon2018} involving only objects of one type, i.e.\ superoperators:
\begin{eqnarray}
 \frac{d}{dt} \supop{U}_S(t) = -i \lvl_S \supop{U}_S(t) - \int_0^t dt' \, \supop{K}(t-t') \supop{U}_S(t'),
\label{eq:system_propagator}
\end{eqnarray}
In this form, it becomes clear that one can solve for $\supop{K}$ directly from $\supop{U}_S$.
Note that no approximations have been made and the dynamics generated by solving \nnref{eq:system_propagator}{(}{)} and \nnref{eq:laplace_memory}{(}{)} are equivalent to solving \nnref{eq:liouville}{(}{)} with the stated assumptions on initial conditions.

The derivation however, benefits from a simplification made possible by the form of the model Hamiltonian in \nnref{eq:model}{(}{)}.
Bath traces over the interaction Liouvillian $\lvl_V$ with respect to a bath state $\op{\rho}_B$ of fixed magnetization will be zero due to the conservation of total magnetization in the model, and if we choose $\op{\rho}_B$ to have zero magnetization.
Therefore the validity of \nnref{eq:laplace_memory}{(}{)} is not restricted to solely thermal baths ($\op{\rho}_B \propto e^{-\beta \op{H}_B}$) nor bath eigenstates ($\comm{\op{\rho}_B}{\op{H}_B} = 0$).

The memory kernel $\supop{K}$ and the system propagator $\supop{U}_S$, being linear mappings from the system Hilbert space $\mathcal{H}_S$ to itself, can be represented as $(\mathrm{dim}\mathcal{H}_S)^2 \times (\mathrm{dim}\mathcal{H}_S)^2$ matrices.
Requirements on unitarity and hermiticity, along with the decoupling of populations and coherences in this magnetization-conserving model, means that $\supop{U}_S$ is described by only two independent entries when the focus is solely on population dynamics.
The same extends to $\supop{K}$ by virtue of its relation to $\supop{U}_S$. 
The two entries of $\supop{U}_S$ are computed by two independent instances of the initial system state $\op{\rho}_{S}(0)$: one from the population of the $|0\rangle$ state when $\op{\rho}_{S}(0) = |0\rangle\langle 0|$, and the other from the population of the $|1\rangle$ state when $\op{\rho}_{S}(0) = |1\rangle\langle 1|$.
The initial bath state is the same in both cases, with definite magnetization $M_B=0$.
Because the total magnetization $\op{M}^z = \op{\tau}^z + \sum_i \spin{z}{i}$ is conserved, these two trajectories must reside in independent parts of Hilbert space.
They are then combined in solving for the memory kernel, which can be done in the time domain by discretizing the integro-differential equation (see the supplementary materials for details).
This, while posing no problem for the projection operator formalism, leads to a strange scenario where the central qubit dynamics restricted to one symmetry sector will depend on information from another, disjoint symmetry sector.

To skirt around this unsavory philosophical scenario, we can focus on only the population of the $|0\rangle$ state of the central qubit.
Using the projection operator $\supop{P} \op{\rho} = (|0\rangle\langle 0| \otimes \op{\rho}_B) \Tr[(|0\rangle\langle 0| \otimes \op{I}_B) \op{\rho}]$, one can repeat the same steps as before and obtain the scalar memory kernel for a single disorder realization as
\begin{eqnarray}
\hspace*{-8pt}  K(t) &= \Tr \left[ (|0\rangle\langle 0| \otimes \op{I}_B) \lvl \supop{Q} e^{-i \supop{Q} \lvl \supop{Q} t} \supop{Q} \lvl (|0\rangle\langle 0| \otimes \op{\rho}_B) \right],
\label{eq:scalar_memory}
\end{eqnarray}
satisfying the integro-differential equation
\begin{eqnarray}
  \frac{d}{dt} p_0(t) = - \int_0^t dt' \, K(t-t') p_0(t'),
  \label{eq:scalar_memory_time}
\end{eqnarray}
or its Laplace-domain equivalent
\begin{eqnarray}
  \widetilde{K}(z) &= -z + \frac{1}{\widetilde{p_0}(z)}.
  \label{eq:scalar_memory_laplace}
\end{eqnarray}
Focusing on the population $p_0(t)$ of single state allows us to work with a scalar memory kernel $K(t)$ and simplifies the calculations.
We will focus exclusively on the scalar memory kernel for the remainder of this paper. 
While this may be an unconventional choice of projector and therefore also of a memory kernel, we stress that the Nakajima-Zwanzig equation in its most general form does \textit{not} depend on the choice of the $\supop{P}$.
The only requirement is that the same observables of interest are contained in the domains of the different projectors.~\footnote{See \cite{Ng2021c} for a detailed demonstration of the equivalence of dynamics generated by different forms of generalized master equations resulting from the interplay of projections and the presence of conserved quantities.}

Note that the memory kernel is akin to the self-energy for the reduced density matrix.
Solving for it is then tantamount to solving the exact problem.
Yet there are still advantages to working with the memory.
For one, because of its relationship with the central qubit's populations it is in principle a measurable quantity.
There is also the possibility for the memory to decay on timescales different from that of the populations.
Should the memory decay much faster, then it may be possible to leverage the timescale separation to reduce the computational effort required to solve for the system dynamics at longer times.

\subsection{Disorder averaged memory}
\label{sec:disordered_memory}
Given that we are interested in disordered systems, suitable definitions of a memory kernel associated with different disorder realizations depends on the quantity of experimental interest.
The difference depends on when the disorder averaging is performed.
We denote by $K_{\mathrm{avg}}$ the case where the population $p_0$ is averaged over the disorder ($\overline{p_0}$) before solving for the memory kernel, satisfying
\begin{eqnarray}
\label{eq:Kavg_convolution}
\frac{d}{dt}\overline{p_0}(t) &= - \int_0^t dt' \, K_{\mathrm{avg}}(t-t') \overline{p_0}(t'). 
\end{eqnarray}
The other case, where the memory for disorder realization is found and then averaged, is denoted by $\overline{K}$.
This latter case is relevant should one decide that the observable of interest is the memory kernel itself, which is in principle possible since it is directly computable from the populations.

It is not \textit{a priori} clear how these two definitions are related.
A reasonable guess might be that, upon disorder averaging, the two definitions are equivalent.
We argue that this is not necessarily correct.
Suppose that for every $L$ the disorder-averaged population $\overline{p_0}(t)$ exists, with initial condition $\overline{p_0}(0) = 1$.
The trajectory of the population for a single instance of disorder will have deviations from this average value, $p_0(t) = \overline{p_0}(t) + \delta p(t)$.
Since populations must be positive at all times, so should their Laplace transforms for real, positive $z$.
Using \nnref{eq:scalar_memory_laplace}{(}{)}, the positivity of the Laplace transforms allows us to write
\begin{eqnarray}
  \widetilde{\delta K}(z) &= \frac{1}{\widetilde{\overline{p_0}}(z) + \widetilde{\delta p}(z)} - \frac{1}{\widetilde{\overline{p_0}}(z)} \nonumber \\
                          &= \int_0^{\infty} du \, e^{-u \left( \widetilde{\overline{p_0}}(z) + \widetilde{\delta p}(z) \right)} - e^{-u \widetilde{\overline{p_0}}(z)} \nonumber \\
  &= \int_0^{\infty} du \, e^{-u \widetilde{\overline{p_0}}(z)} \left( e^{-u \widetilde{\delta p}(z)} - 1 \right).
\end{eqnarray}
Since the exponential function is entire, the term in parentheses can be expanded as a series,
\begin{eqnarray}
\widetilde{\delta K}(z) &= \int_0^{\infty} du \, e^{-u \widetilde{\overline{p_0}}(z)} \sum_{n=1}^{\infty} \frac{(-u)^n}{n!} \left( \widetilde{\delta p}(z) \right)^n.
\label{eq:memory_flucs}
\end{eqnarray}
Averaging this expression over disorder, we will have the $n=1$ term vanish by definition of $\delta p$.
But all higher order terms--particularly ones with even powers--are not guaranteed to vanish.
The consequence is that $K_{\mathrm{avg}} \neq \overline{K}$ for finite $L$.

\begin{figure*}
\centering
  \includegraphics[width=\textwidth]{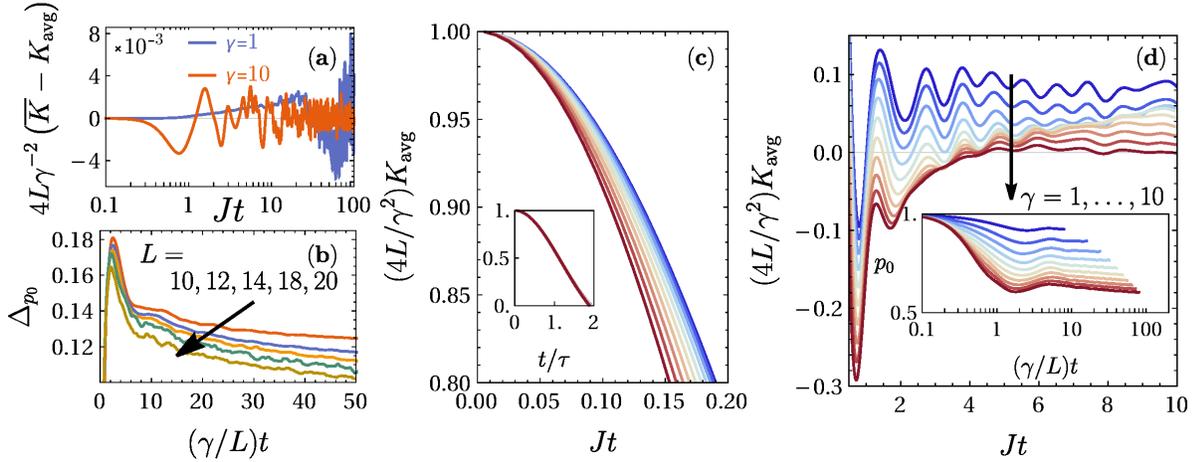}
  \caption{\textbf{(a)} Comparison of the averaged memory kernel $\overline{K}$ and the memory kernel of the average $K_{\mathrm{avg}}$ for $L=16$ deeply in the localizing ($\gamma=1$) and thermalizing ($\gamma=10$) phases, with $\geq 500$ disorder realizations. \textbf{(b)} Root-mean-squared fluctuations $\Delta_{p_0}$ of the population for $\gamma=10$. \textbf{(c,d)} The scalar memory kernel of the averaged population, $K_{\mathrm{avg}}(t)$, for $L=12$, onsite disorder strength $W=6.0$, and with 6400 disorder realizations. The memory is rescaled such that its initial value is $1$, and separated into the \textbf{(c)} short and \textbf{(d)} intermediate time regimes. For clarity, the data in the main panel of \textbf{(d)} are shifted up in multiples of $0.1$ away from the $\gamma=10$ curve. \textbf{(c, inset)} Collapse of the short time memory upon rescaling the time by $\tau$ defined in \nnref{eq:shortTimeExpansion}{(}{)}. \textbf{(d, inset)} The populations of the $|0\rangle$ state for the central qubit used to generate $K_{\mathrm{avg}}$.}
  \label{fig:memory_split}
\end{figure*}

The situation is modified in the thermodynamic limit owing to self-averaging.
Intuitively, a small subsystem interacting randomly with $N\gg 1$ degrees of freedom should have deviations from its mean behavior that decrease as $N$ increases.
As a result, when the environment is sufficiently large, a single realization of the random interaction should typically yield results close to the mean.
This statement was recently demonstrated~\cite{Ithier2017}, showing that the system reduced density matrix enjoys the typicality property for system-bath interactions modeled by certain classes of random matrices.
Importantly, \cite{Ithier2017} showed that this self-averaging property holds at least up to a timescale $T$ that increases with $L$.
Thus if $p_0(0 \leq t \leq T)$ is self-averaging, so must $K(t)$ on the same interval, since to solve for $K(t)$ up to time $T$ in \nnref{eq:scalar_memory_time}{(}{)} requires only $p_0(t)$ on $[0,T]$.
In \nnref{fig:memory_split}{figure~}{b} we show the root-mean-squared fluctuations of $p_0(t)$ deeply in the thermalizing phase of the bath-disordered Hamiltonian \nnref{eq:model}{(}{)}, and observe that they indeed decrease with increasing bath size.
Extrapolating to the thermodynamic limit, we should therefore have self-averaging of the reduced density matrix of the central qubit.
Then by extension the memory must self-average too.
This can be seen from \nnref{eq:memory_flucs}{(}{)}, where fluctuations of a single realization of $K(t)$ has deviations from $K_{\mathrm{avg}}(t)$ that are bounded by the magnitude of the fluctuations in the population $\delta p(t) = p_0(t) - \overline{p_0}(t)$.
In \nnref{fig:memory_split}{figure~}{a}, we find that $\overline{K}(t)$ and $K_{\mathrm{avg}}(t)$ generally tend to differ by $|\overline{K}(t) - K_{\mathrm{avg}}(t)| \sim O(10^{-3} (\gamma^{2}/4L))$ up to timescales $t \lesssim O(10^2)$ for the system sizes we can simulate.
We observe that this deviation can diverge exponentially with a finite number of disorder realizations at long enough times, a phenomenon which we will return to in \nnref{sec:long_times}{section~}{}.
Barring that, the self-averageness of the population $p_0(t)$--which yields $K_{\mathrm{avg}}(t) = K(t) \Longrightarrow K_{\mathrm{avg}}(t) = \overline{K}(t)$ in the thermodynamic limit--gives us an alternate window into understanding how the memory kernel behaves.
For the remainder of this paper, we shall mostly discuss $K_{\mathrm{avg}}(t)$ as we are interested also in the dynamics of the averaged population.

\section{Results}
\label{sec:results}
We implement time evolution by approximating $e^{-i \op{H} t}$ with Chebyshev polynomials~\cite{Kosloff1994, Weisse2006}.
To reduce computational costs, we use the conservation of total magnetization $\op{M}^z = \op{\tau}^z + \sum_i \spin{z}{i}$ in the model, allowing us to restrict the dynamics to the symmetry sector with $\op{M}^z = -1$.
The system is prepared in the $\op{\rho}_{S,0} = |0\rangle\langle 0|$ state, while the bath state $\op{\rho}_B$ is initialized to be a Neel state, $|\cdots\downarrow\uparrow\downarrow\uparrow\downarrow\cdots\rangle$.
We expect similar results should we choose different initial states within the sector of $\op{M}^z = -1$.

A (matrix) memory kernel $\supop{K}$ with $n$ independent entries can be computed directly from the populations~\cite{Kidon2018} using $n$ different initial conditions, for each disorder realization.
In this sense, there is added computational benefit to restricting our discussion to only the scalar memory kernel $K(t)$.

\subsection{Short times}
\label{sec:short_times}
We can leverage the self-averaging property to gain some understanding of the short time behavior (\nnref{fig:memory_split}{figure~}{c}) of $K_{\mathrm{avg}}(t)$.
The derivatives of $K(t)$ at $t=0$ for a single disorder realization can be found straightforwardly (see the supplementary materials) from those of $p_0(t)$, with the lowest orders being
\begin{eqnarray}
  K(t=0) &= -p_0^{(2)}(t=0) \nonumber \\
  K^{(2)}(t=0) &= - p_0^{(4)}(t=0) + \left( p_0^{(2)}(t=0) \right)^2,
\end{eqnarray}
where $f^{(n)}$ denotes the $n$-th derivative.
After averaging over disorder with an initial Neel state in the bath, we have
\begin{eqnarray}
  \frac{K_{\mathrm{avg}}(t)}{\gamma^2/4L} &\approx 1 - \frac{1}{2} \left( \frac{t}{\tau_K} \right)^2 + O(t^4) \nonumber \\
  \frac{1}{\tau_K} &\equiv \sqrt{\frac{W^2}{3} + \frac{3J}{4} \frac{\gamma}{L} + \frac{3}{4} \frac{\gamma^2}{L} - \frac{3}{4} \frac{\gamma^2}{L^2}},
  \label{eq:shortTimeExpansion}
\end{eqnarray}
where $J = 1$ in our model, and $L$ is the number of spins in the bath.
We see that the disorder strength $W$ sets the initial decay rate $1/\tau_K$.
This can be roughly estimated for large $W$ from Fermi's Golden Rule, using our argument that disorder-averaging effectively gives us a continuous spectrum with an effective (root-mean-squared) bandwith $\sim O(W\sqrt{L})$, and a coupling strength $\sim (\gamma/2 L)^2$.
While $W$ sets the decay timescale for $K_{\mathrm{avg}}(t)$, it is the quantity $\gamma^2/4 L$ that sets the overall magnitude of $K_{\mathrm{avg}}(t)$ and so dictates the timescale for $p_0(t)$.
We expect so from the following scaling argument:
Assume that the memory kernel converges to a limiting form in the thermodynamic limit as
\begin{eqnarray}
\lim_{L\to\infty} \frac{K_{\mathrm{avg}}(t)}{\gamma^2/4L} = k(t),
\end{eqnarray}
where $k(t)$ is independent of $L$ and has a short time expansion given by \nnref{eq:shortTimeExpansion}{(}{)}.
From the Nakajima-Zwanzig equation,
\begin{eqnarray}
\frac{d p_0}{dt} \approx - \frac{\gamma^2}{4L} \int_0^t d\tau \, k(\tau) p_0(t - \tau),
\end{eqnarray}
we rescale the time to $t' = (\gamma^r/L^s) t$ and obtain
\begin{eqnarray}
  \frac{d p'_0}{dt'} = - \frac{\gamma^{2-2r} L^{2s-1}}{4} \int_0^{t'} d\tau' \, k\left( \frac{L^s \tau'}{\gamma^r} \right) p'_0(t' - \tau'),
  \label{eq:rescaledNZ}
\end{eqnarray}
where $p'_0(t') \equiv p_0(t'/(\gamma^r L^{-s}))$.
We seek exponents $r>0$ and $s>0$ such that $p'_0(t')$ will vary on the timescale $\Delta t' \sim 1$.
With the rescaled time, the $k(L^s \tau'/\gamma^r)$ appearing in \nnref{eq:rescaledNZ}{(}{)} will have largely decayed by $\tau' \sim \gamma^r L^{-s}/(W/\sqrt{3})$, a timescale much faster than that of $p'_0(t')$.
Hence we can approximate $p'_0(t'-\tau')$ in \nnref{eq:rescaledNZ}{(}{)} as a constant, and estimate the strength of memory effects by integrating $k(L^s \tau'/\gamma^r)$ up to its decay time.
This is roughly given by
\begin{eqnarray}
\left( \frac{\gamma^{2-2r} L^{2s-1}}{4} \right) \left( \frac{\gamma^r/L^s}{W/\sqrt{3}} \right) = \frac{\sqrt{3}}{4} \frac{\gamma^{2-r}}{W} L^{s-1}.
\end{eqnarray}
We require $s = 1$ in order to have a converged $p'_0$ on the timescale of $t'$ in the thermodynamic limit.
Furthermore, $r = 1$ so that a trivial rescaling of the Hamiltonian $\op{H} \to \alpha \op{H}$ would not alter the strength of the memory term.
Thus we argue that the dynamics of the central qubit should proceed on the timescale $\tau_{p_0} \sim L/\gamma$, consistent with our initial assumption that the population dynamics proceed much more slowly than does its associated memory kernel.
We show this rescaling of time in \nnref{fig:memory_split}{figure~}{b} and in the inset of \nnref{fig:memory_split}{figure~}{d}, where the former shows the fluctuations of $p_0(t)$ between disorder realizations for different system sizes at fixed $\gamma=10$, and the latter shows $p_0(t)$ for fixed $L=12$ across $\gamma$.
These figures show that the lowest moments of the populations align on the timescale $\tau_{p_0} \sim L/\gamma$.
This result is also consistent with the result of \cite{Hetterich2018} on the central qubit's autocorrelation function, $\int d\tau \lrangle{\op{\tau}^z(t+\tau)}\lrangle{\op{\tau}^z(\tau)}$, where it was observed that there is an accumulation of spectral weight near $\omega \sim \gamma/L$.

With a clear separation between $\tau_K$ and $\tau_{p_0}$, one may wonder whether the central qubit can be described by an effective master equation.
At least deep in the localized phase, the bath is too slow to act as an effective reservoir for the central system.
Correlation functions of the bath are argued~\cite{Gopalakrishnan2015, Gopalakrishnan2020} to decay as a power law $t^{-\zeta}$ with $0 < \zeta < 1$, which makes memory effects crucial in dictating the behavior of $p_0(t)$ at long times.
We will return to discuss the long time behavior of the memory kernel below in \nnref{sec:long_times}{section~}{}.

\subsection{Intermediate times}
\label{sec:intermediate_times}
As seen in \nnref{fig:memory_split}{figure~}{d}, the memory past $J t\gtrsim 1$ takes on different behaviors depending on the coupling strength, with increasingly damped oscillations as the combined system and bath transitions from localization to thermalization.
The inset of \nnref{fig:memory_split}{figure~}{d} shows that this behavior is not observable when looking solely at the populations.
The oscillation is dominated by frequencies in the range $\omega \in (4,6)$, close to the disorder strength $W = 6$.
Such oscillations are not a feature unique to an interacting bath.
They show up in the non-interacting limit $J=0$, in which the memory to lowest order in $\gamma$ can be approximated by
\begin{eqnarray}
K_{J=0}(t) &\approx \frac{\gamma_{\perp}^2}{L} \frac{\sin \left( W t \right)}{W t} + O(\gamma^3),
\end{eqnarray}
where $\gamma_{\perp} = \gamma/2$.
We see that oscillations are linked to the finite bandwidth $W$ of frequencies in the bath~\cite{Cohen2011}, which arises from precession about the local field on each site, $(h_i/2) \spin{z}{i}$, and $h_i \in [-W,W]$.
When interactions in the bath are turned on, we would expect them to provide a small renormalization to the precession frequencies, as we are working with a hierarchy of scales such that $W \gg J > \gamma$.
This assumes, of course, that the bath dynamics are approximately describable with a precession picture even in the presence of bath interactions.

To justify this picture more formally, we can leverage the description of MBL systems in terms of quasi-local integrals of motion, which form the effective bath degrees of freedom that exhibit precession.
At intermediate times and at weak coupling, the memory kernel can be approximated by bath correlation functions~\cite{Breuer2002,Nitzan2006},
\begin{eqnarray}
K(t) \approx \frac{\gamma_{\perp}^2}{L^2} \sum_{i,j,\pm} \Tr \left[ \spin{\pm}{i}(t) \spin{\mp}{j}(0) \op{\rho}_B \right].
\label{eq:memory_bath_corrfxs}
\end{eqnarray}
In the MBL phase, the bath spin operators $\spin{\pm}{i}$ have large overlaps~\cite{Huse2014} with quasi-local operators $\op{\Theta}^{x,y,z}_i$ with which the bath Hamiltonian can be written as~\cite{Huse2014,Imbrie2017,Gopalakrishnan2020} 
\begin{eqnarray}
  \op{H}_B = \sum_{i=1}^L \varepsilon_i \op{\Theta}^z_i + \sum_{i,j} J_{i,j}\op{\Theta}^z_i \op{\Theta}^z_j + \cdots,
  \label{eq:liom_hamiltonian}
\end{eqnarray}
where the operators $\op{\Theta}^{x,y,z}_i$ follow the Pauli commutation relations.
The bath correlation functions oscillate according to $\varepsilon_i$, at least when the bath is strongly localized.
The distribution of $\varepsilon_i$ will therefore dictate the intermediate-time behavior of the memory kernel.
For instance, if the distribution has sharp cutoffs like in the case of box disorder, then it can be expected that the memory will display oscillatory behavior whenever the stated approximations are applicable.
We note that the picture of precessions is complicated at later times by dephasing mechanisms arising from interactions--the 2-body $J_{i,j}$ terms and higher--in the bath.
Therefore, measurement of the memory kernel will yield some information on the parameters entering the bath Hamiltonian \nnref{eq:liom_hamiltonian}{(}{)}.

At the other extreme, where the system strongly couples ($\gamma \gtrsim 5$ for the value of $W=6$ we have shown in \nnref{fig:memory_split}{figure~}{d}) to the bath, the localization assumed above breaks down~\cite{Ponte2017}.
That is, the bath interactions mediated by the qubit are strong enough that the bath cannot remain ``close'' to its initial state, so the expansion of the memory in terms of bath correlation functions no longer holds.
In all, the contribution of the bath to the system dynamics can no longer be parsed into contributions from (nearly) independent oscillators.
Instead, delocalization evidently serves to homogenize the influence of the bath, smoothing over the randomness from the local fields $h_i$, and damping out oscillations in $K_{\mathrm{avg}}(t)$ as observed in the red curves of \nnref{fig:memory_split}{figure~}{d}.

\begin{figure}
  \centering
\includegraphics[scale=0.9]{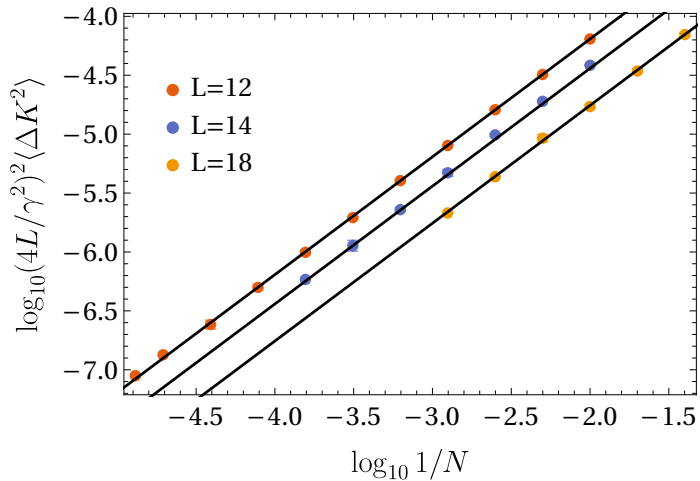}
  \caption{Average squared deviations on $t\in [50, 100]$ of $K_{\mathrm{avg}}(t)$ in the thermalizing regime ($\gamma = 10$), as a function of disorder realizations $N$. Solid lines denote $1/N$ decay and serve as guides to the eye.}
  \label{fig:tailFlucs}
\end{figure}

\subsection{Long times}
\label{sec:long_times}
Within the particular parameters we have chosen to study in this model, we define ``long times'' to correspond to $Jt \gtrsim 10$, a time past which the coherent oscillations in the bath have dephased. 
For the purpose of extrapolating the dynamics, it is crucial to understand how quickly $K_{\mathrm{avg}}(t)$ decays, if it even does so at all.
However, since we can only numerically average over a finite number $N$ of disorder realizations, we cannot expect to observe a clear decay signal.
Instead, we can ask whether the long time behavior of $K_{\mathrm{avg}}(t)$ is consistent with small, possibly vanishing, values should we extrapolate our results to infinite $N$.
Deeply in the thermalizing phase, we show in \nnref{fig:tailFlucs}{figure~}{} that the magnitude of time-averaged fluctuations
\begin{eqnarray}
  \lrangle{\Delta K^2}_{[T_i, T_f]} = \lrangle{K^2_{\mathrm{avg}}(t)}_{[T_i, T_f]} - \lrangle{K_{\mathrm{avg}}(t)}^2_{[T_i, T_f]},
  \label{eq:sqDev}
\end{eqnarray}
in the tail portion $K_{\mathrm{avg}}(50 \leq t \leq 100)$ decays as $1/\sqrt{N}$, and moreover decreases with increasing system size as would be expected from self-averaging systems.
In the above equation, we use the notation $\lrangle{g(t)}_{[T_i, T_f]} = \int_{T_i}^{T_f} dt \, g(t) / (T_f - T_i)$.

The persistence of the finite $N$ noise makes it difficult to conclusively show numerically whether $K(t)$ decays as algebraically or exponentially. 
While in \nnref{sec:short_times}{section~}{} we argued for a power law decay for the weakly coupled, localized phase based on known phenomenology of MBL, this approach cannot work for the strongly coupled, thermalizing phase.
In the absence of weak coupling perturbative expansions we now turn to the self-averaging relations $K_{\mathrm{avg}} \sim K \sim \overline{K}$ to attempt to extract insights about the thermalizing phase.
Doing so requires discussion about the memory kernel for a single realization of disorder, which is what we shall focus on for the remainder of this subsection.

For certain realizations of $\{h_i\}$, we observe an increasing likelihood for the memory--both scalar- and matrix-valued versions--to display unbounded exponential divergences with increasing coupling $\gamma$.
We can verify the divergence for small system sizes $L\lesssim 6$, where the Laplace transformed memory kernel can be computed directly to yield the memory as a sum over simple poles, some of which with positive real parts.
Such contributions--which are necessary in order to correctly reproduce the population dynamics--lead to an unbounded exponential \textit{increase} of the memory for particular values of the coupling and magnitude of disordered fields.
We will return to discuss the origins and implications of such pathological behavior in \nnref{sec:exp_growth}{section~}{}.

We can motivate the consequences of exponentially growing contributions to $K(t)$ by examining the structure of the poles of its Laplace transform, $\widetilde{K}(z)$.
Because the Hamiltonian is real and Hermitian, poles of $\widetilde{K}(z)$ are given by a real polynomial (see the supplementary materials for details).
The polynomial will only involve terms of even powers, $z^{2n}$, because $p_0(t) = p_0(-t)$.
Thus if a pole $s_n$ exists with residue $r_n$ such that $\mathrm{Re}\, s_n \neq 0$, it must be the case that poles $-s_n$, $s^{*}_n$ and $-s^{*}_n$ must exist with residues $r_n$, $r_n^{*}$, and $r_n^{*}$ respectively.
Based on the distribution of the pole structure, any exponentially dampened part of the memory ($\mathrm{Re} \, s_n < 0$) must be accompanied by an exponentially growing counterpart.
We posit that in the thermodynamic limit one of two situations must hold: 1) all off-axis poles converge towards the $\mathrm{Im}\, z$ axis as $L\to\infty$, or 2) some poles still exist off-axis, which because of the conjugate pairs, contributes both exponential decay and growth.
In the first scenario, there are no isolated poles to cause exponential decay.
In the second scenario, any exponential decay is masked by exponential growth. Moreover, even if $\mathrm{Re}\, s_n > 0$ poles cancel upon disorder averaging, the same would happen to the $\mathrm{Re}\, s_n < 0$ poles by virtue of the relationship between residues discussed above. 
Therefore we argue that even in the thermalizing phase, the memory kernel for the dynamics we have defined should not exhibit exponential decay in the limit as $L\to\infty$.
This leaves open the possibility of power-law or stretched-exponential behavior.
In the next section, we will use infinite-time data from exact diagonalization to show that the long-time behavior of the memory is consistent with a power-law decay.
Finally, we reiterate the importance of the order of limits in this problem.
They must be taken as
\begin{eqnarray}
\lim_{t\to\infty} \lim_{L\to\infty} \lim_{N\to\infty}
\end{eqnarray}
to ensure that, reading from right to left, the population--and therefore the memory kernel--does not recur and to ensure the validity of the approximation $\overline{K} \approx K_{\mathrm{avg}}$.

\section{Extracting long time information from the memory kernel}
\label{sec:extraction}
The memory kernel has a direct relation to steady state values of the reduced density matrix, provided that a steady state exists~\cite{Cohen2011, Cohen2013, Wilner2013, Wilner2014}.
While the past work was done using all $d^2 \times d^2$ elements of the (matrix) memory kernel, we can import their ideas to the scalar memory kernel and a single element of the reduced density matrix.
From the relationship between $\widetilde{\overline{p_0}}(z)$ and $\widetilde{K}_{\mathrm{avg}}(z)$, we can use the final value theorem to find
\begin{eqnarray}
  \label{eq:finalValueTheorem}
  \lim_{z\to 0} z \widetilde{\overline{p_0}}(z) &= \lim_{z\to 0} \frac{1}{1 + \widetilde{K}_{\mathrm{avg}}(z)/z} \\[-15pt]
 \lim_{t\to\infty} p_0(t) &= \lim_{z\to 0} \bigg[ 1 + \int_0^{\infty} dt \, e^{-z t} \overbrace{ \int_0^t d\tau \, K_{\mathrm{avg}}(\tau)}^{\equiv \kappa(t)}\bigg]^{-1}. \nonumber
\end{eqnarray}
If $\kappa(t)$ decays sufficiently quickly, we can extrapolate the $z\to 0$ limit from the finite times accessible from numerics.
However, as we argued in the previous section, the memory cannot decay exponentially; therefore there is no single cutoff time $t_c$ that can be used to approximate
\begin{eqnarray}
\lim_{z\to 0} \int_0^{\infty} dt \, e^{-z t} \int_0^t d\tau \, K_{\mathrm{avg}}(\tau) &\approx \int_0^{t_c} dt \, \int_0^t d\tau \, K_{\mathrm{avg}}(\tau).
\end{eqnarray}
A long time tail of $K_{\mathrm{avg}}(t)$ would have non-negligible contributions to the dynamics, and therefore much care has to be taken in its use for extrapolations.

\begin{figure*}
\centering
  \includegraphics[width=\textwidth]{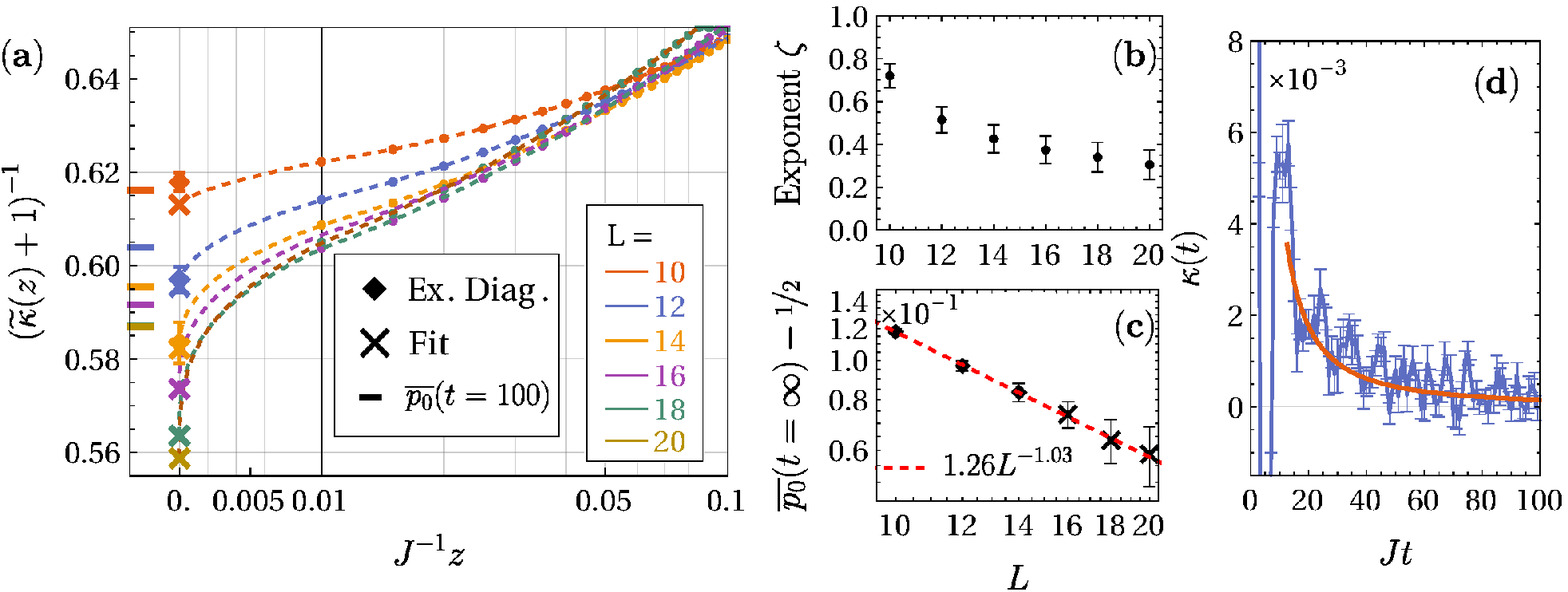}
  \caption{\textbf{(a)} Infinite time value of the average population $\overline{p_0}$ deeply in the thermalizing phase ($\gamma = 10$) using \nnref{eq:finalValueTheorem}{(}{)} and \nnref{eq:powerLawAnsatz}{(}{)}. Dashed lines and black crosses indicate respectively the fit to \nnref{eq:powerLawAnsatz}{(}{)} and the extrapolated value for $t_c = \infty$. Where available, squares indicate the long time ($t\sim 10^{12}$) value of $p_0$ calculated independently from exact diagonalization. \textbf{(b)} Fitted exponents $\zeta$ as a function of system size. \textbf{(c)} Log-log plot of the long time limit of $\overline{p_0}$ versus the number of sites in the bath, $L$. Error bars in \textbf{(b,c)} of the extrapolated quantities ($L\geq 16)$ correspond to 95\% confidence intervals for the parameter estimation. \textbf{(d)} The integrated memory $\kappa(t)$ for $L=12$ over $7.7\times 10^5$ realizations of disorder. The red curve is the asymptotic time-domain behavior of $z^{\zeta} \Longrightarrow t^{-1-\zeta}/\Gamma(-\zeta)$, as extracted from the fit to \nnref{eq:powerLawAnsatz}{(}{)}.}
  \label{fig:yintsFit}
\end{figure*}

In lieu of a cutoff approximation, we turn again to the definition of $K_{\mathrm{avg}}$,
\begin{eqnarray}
  \widetilde{K}_{\mathrm{avg}}(z) &= -z + \frac{1}{\widetilde{\overline{p_0}}(z)}.
  \label{eq:kavg_laplace}
\end{eqnarray}
We take an ansatz for the memory at small $z$,
\begin{eqnarray}
\widetilde{\kappa}(z) \equiv \frac{\widetilde{K}_{\mathrm{avg}}(z)}{z} &\approx \left( -1 + \frac{1}{p^{\infty}} \right) + a_0 z^{\zeta} + a_1 z,
  \label{eq:powerLawAnsatz}
\end{eqnarray}
where $0 < \zeta < 1$ and the long time limit of the average population $\overline{p_0}(t)$ shall be denoted as $p^{\infty}$.
Note that we had argued in the previous section at least for the \textit{absence} of exponential decay of the memory kernel in the thermodynamic limit, based on the structure of the poles in Laplace space.
The presence of terms like $z^\zeta$ is consistent with long-time behavior as $\kappa(t) \sim t^{-\zeta-1} \Longrightarrow K_{\mathrm{avg}}(t) \sim t^{-\zeta-2}$.

To extrapolate the long time populations, we compute $\kappa(t)$ defined in \nnref{eq:finalValueTheorem}{(}{)} and approximate its Laplace transform 
\begin{eqnarray}
\widetilde{\kappa}(z) &\approx \int_0^{t_{\mathrm{max}}} dt \, e^{-z t} \kappa(t).
\end{eqnarray}
This result is then fitted using \nnref{eq:powerLawAnsatz}{(}{)} to find $p^{\infty}$ and $b_0$ and the amplitudes $a_n$.
Such an approximation for the Laplace transform is admissible only if $\kappa(t)$ has decayed to sufficiently small values at $t = t_{\mathrm{max}}$, and for $z \gtrsim t_{\mathrm{max}}^{-1}$.
We find that the results of using such an extrapolation procedure agree well with the values from independent calculations using exact diagonalization (\nnref{fig:yintsFit}{figure~}{a}).
Thus we are able to obtain estimates for the long-time population of the central qubit for system sizes ($L\gtrsim 16$) larger than those obtainable through exact diagonalization.
In particular, this allows us to see how the central qubit approaches the thermalized limit $p_0 = 1/2$ with increasing bath size.
In \nnref{fig:yintsFit}{figure~}{c}, $p^{\infty}$ is consistent with power law decay $p^{\infty} - 1/2 \sim L^{-1.03}$, which is in line with the scaling given by the infinite temperature phase space average,
\begin{eqnarray}
  \frac{\mathcal{H}_{|0\rangle}(M^z = -1)}{\mathcal{H}(M^z = -1)} = \frac{\binom{L}{L/2}}{\binom{L+1}{L/2}} = \frac{1}{2} + \frac{1}{2(L+1)},
\end{eqnarray}
measuring the relative sizes of the Hilbert spaces for eigenstates occupying $|0\rangle$ and $|1\rangle$.
We stress that because the memory must decay with time, this procedure cannot be used in finite systems for a single disorder realization, as the population will generally not reach a steady state in such circumstances.

Furthermore, since we have estimates of the true value of $p^{\infty}$ obtained independently from exact diagonalization, we can compare \nnref{eq:powerLawAnsatz}{(}{)} to a more generic alternative where $\widetilde{\kappa}(z)$ is analytic about $z=0$.
Such is the case if $\kappa(t)$ were to, for example, decay exponentially or faster.
For different system sizes and disorder distributions, we have found that only the power-law ansatz is able to smoothly interpolate between known $z=0$ values of $\widetilde{\kappa}(z)$ from exact diagonalization and $z>0$ values of $\widetilde{\kappa}(z)$ calculated from finite time dynamics (see the supplementary materials for an example).
Thus, while we have been unable to mathematically prove the existence of a long-tail in $K_{\mathrm{avg}}(t)$, we have at least found numerical corroboration for the validity of our claim.

We note that, at least for $L \leq 14$, we find that the exponent $\zeta$ is system size dependent, for both box (\nnref{fig:yintsFit}{figure~}{b}) and Gaussian distributed disorder.
In the absence of intrabath interactions ($J=0$), we observe that the integrated kernel $\kappa(t)$ acquires a large oscillatory component with a decaying envelope at long times for $\gamma=10$, which dominates over the $\kappa(t) \sim t^{-1-\zeta}$ behavior seen with $J=1$ (cf.\ \nnref{fig:yintsFit}{figure~}{d}), see section four of the supplementary materials.
We further argue in the supplementary materials that if one takes the bath to initially be at infinite temperature, there will be a temporal power-law decay of the memory as $\sim t^{-3}$ which implies that $\zeta\to 1$ in this limit.
Altogether, this suggests that $\zeta$ is at least a quantity dependent on intrabath interactions as well as the initial state; we cannot clarify whether there is a limiting value as $L\to\infty$ for initial states of fixed energy density, such as that considered in this work.

One may wonder what advantage this method confers to obtaining infinite-time populations, compared to simply simulating the population dynamics to longer time.
For one, it is not always clear the timescales at which one can be sure that the system will have relaxed.
This point is made more salient by the possibility of small, long-tailed memories which implies similar behaviors in the population dynamics.
In this work, we have argued that it suffices to be able to observe whether the memory has reached the regime of power-law decay, at which point one can use \nnref{eq:powerLawAnsatz}{(}{)}.
We stress that power-law behavior may become more apparent at earlier times in the memory than compared to the population, such as what we have observed in this work.
While the $t^{-\zeta-1}$ contribution to $\kappa(t)$ may be subtle--on the order of $10^{-3}$ in all system sizes and disorder distributions we examined (see \nnref{fig:yintsFit}{figure~}{d} for an example)--it is always possible to systematically improve its resolution simply by performing more disorder averaging.

\begin{figure*}
  \centering
  \includegraphics[width=0.8\textwidth]{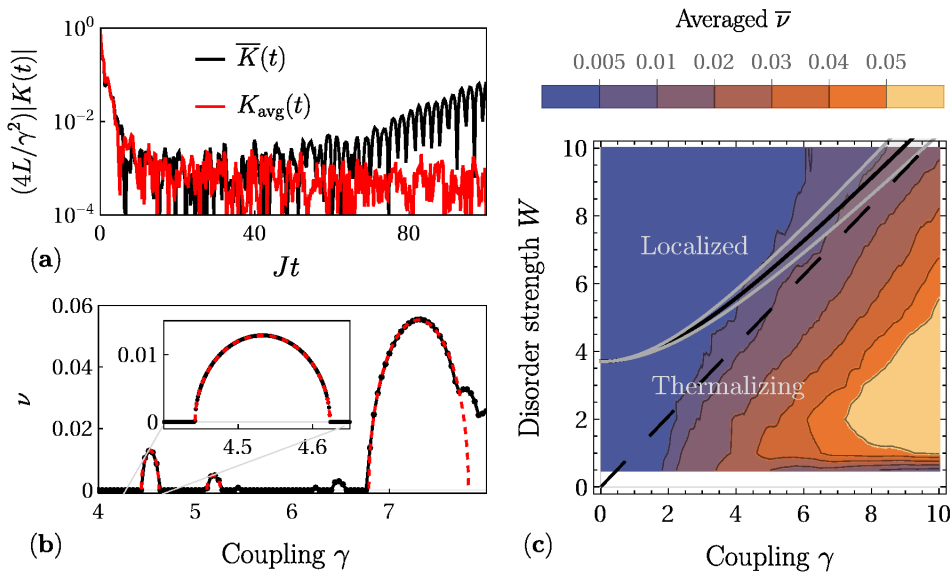}
  \caption{\textbf{(a)} The averaged memory kernel $\overline{K}$ and the memory kernel of the averaged dynamics $K_{\mathrm{avg}}$, for $L=14$ and $\gamma = 10$. The two curves are approximately the same up to $t \lesssim 40$, past which they diverge exponentially owing to certain disorder realizations contributing to $\overline{K}$. \textbf{(b)} Maximum rate $\nu$ of exponential growth for $L=4$ across a range of couplings with a fixed realization of disorder. The rate is computed by solving for the poles of Laplace-transformed memory kernel using 4096 bits of precision. Transitions from zero $\nu$ to finite $\nu$ are sharply discontinuous, and are well captured by fits to half ellipses (dashed red lines). \textbf{(c)} Disorder averaged $\overline{\nu}$ for $L=4$. The black line and its surrounding error bands indicate the $L\to\infty$ phase boundary determined in Ref.~\cite{Hetterich2018}. The dashed black line is the asymptotic behavior of the boundary as argued in Ref.~\cite{Ponte2017}.}
  \label{fig:maxRe}
\end{figure*}

\section{Unbounded exponential growth of the memory kernel}
\label{sec:exp_growth}
We return now to the observation made in \nnref{sec:long_times}{section~}{} about memory kernels growing exponentially in time for certain realizations of the disorder.
As seen in \nnref{fig:maxRe}{figure~}{a}, this can show up in the disorder averaged memory $\overline{K}(t)$, which can only be approximated via sampling over a finite number of disorder realizations.
In \nnref{fig:maxRe}{figure~}{b} we show the maximum real part of the poles--corresponding to the maximum rate of exponential growth $\nu$--for specific set of $\{h_i\}$ with $L=4$.
Intriguingly, $\nu$ is not monotonic with respect to $\gamma$, and displays square root singularities when going from $\nu=0$ to finite $\nu$.
The sharpness of these singularities even with $L=4$ indicates that they should not be associated with thermodynamic phase transitions.
Instead, we believe they stem from exceptional points (EPs) in the generator of projected dynamics, $\supop{Q} \lvl \supop{Q}$, which are related to generalized avoided crossings.
This generator is responsible for the time evolution of the memory kernel, as seen in \nnref{eq:laplace_memory}{(}{)}.
By choosing to focus on only a subset of all the physical degrees of freedom in the problem, we were forced to define projection operators $\supop{P}$ that are not self-adjoint in the space of operators~\cite{Wilkie2001a, Wilkie2001b}.
For example, in operator space the projection operator associated with the scalar memory kernel is $\supop{P} = \big| |0\rangle\langle 0| \otimes \op{\rho}_B \big) \big( |0\rangle\langle 0| \otimes \op{I}_B \big|$, where the adjoint of the operator state vector has action
\begin{eqnarray}
\big( \op{A} \big| \op{B} \big) &= \Tr \left( \op{A}^{\dagger} \op{B} \right).
\end{eqnarray}
The condition of being self-adjoint Liouville space is
\begin{eqnarray}
\supop{P}^{\dagger} = \left( \sum_i \big| \op{A}_i \big) \big( \op{B}_i \big| \right)^{\dagger} = \sum_i \big| \op{B}_i \big) \big( \op{A}_i \big| = \supop{P}.
\end{eqnarray}
Writing $\supop{P}$ in this way, it is clear that even if we project on to a thermal state of the bath, $\op{\rho}_B \propto e^{-\beta H_B}$, the projector $\supop{P}$ still cannot be self-adjoint unless the bath is in an infinite temperature state.
Thus the projected Liouvillian $\supop{Q} \lvl \supop{Q}$ is also not self-adjoint, a property which allows EPs to occur.
We have verified that the same phenomenon occurs even if we work with larger projection superoperators leading to matrix-valued memory kernels.
We have additionally verified numerically that features unique to EPs such as the coalescence of eigenvalues and self-orthogonality are also present (see supplementary materials).

Interestingly, we note that the region in $(W, \gamma)$-space (\nnref{fig:maxRe}{figure~}{c}) for which MBL is predicted to be stable in the thermodynamic limit appears to be correlated with a suppressed $\nu$.
While we are currently unable to prove that this is not a coincidence--such system sizes cannot inform us about the stability of MBL
-- it is possible that this provides a window into the character of the eigenstates, which are argued to be radically altered at large enough $\gamma$ due to percolating networks of resonance states~\cite{Ponte2017}.
At the same time, it is known that the presence of exceptional points limits the radius of convergence for perturbative expansions~\cite{Heiss2012,Marie2021}, and is postulated to be linked to quantum phase transitions~\cite{Jung1999, Garmon2012}.
To fully explore any link between exceptional points, delocalization, and the breakdown of perturbative approaches to MBL will require a separate, in-depth study.

Heuristically speaking, delocalization with increasing coupling is the result of singular behavior in the full Hamiltonian, a fact which should be reflected in both the eigenstates and the spectrum.
In finite systems, these may be isolated occurrences whose singular properties are smoothed out upon taking expectation values.
Our numerical observations suggest that the non-Hermiticity of the projected Liouvillian is highly sensitive to such singularities.
We suspect this may be further indication of a deeper connection between localization and long-time pathologies in the memory kernel, but we are unable to clarify the underlying physics at this time.
However, we will note that the situation may be altered by introducing a large bias on the central qubit, e.g.\ $\Omega \op{\tau}^z$, the analysis of which we will leave for future work.

\section{Discussion and conclusions}
\label{sec:discussion}
In this work we have undertaken the study of the time-nonlocal memory kernel describing how a many-body localizable ``bath'' affects the population dynamics of a central qubit.
While the memory is formally defined in terms of Liouvillians, the dimensions of which quickly grow to be computationally intractable with increasing size of the Hilbert space, we are able to compute it numerically exactly from existing methods for simulating dynamics in closed quantum systems~\cite{Kidon2018}.
We note in passing that the method we use in this work is general, and can easily be formulated to describe the dynamics of the central qubit's coherence, as might be relevant for some recent NMR experiments~\cite{Niknam2020}.
With this method we are able to directly examine the behavior of the memory kernel, parsing it into three regimes: short, intermediate, and long times.

On short timescales ($Jt \lesssim 1$)
is where the majority of the memory's decay occurs, irrespective of whether localization (at small $\gamma$) or delocalization (at large $\gamma$) is present.
Properties of the memory on this timescale largely dictate the timescale of the dynamics for the central qubit's populations.
We note that \nnref{eq:shortTimeExpansion}{(}{)} holds for arbitrary unbiased (i.e.\ zero mean) distributions of onsite fields with variance $W^2/3$ and bath-bath interaction strength $J$.
On intermediate timescales ($Jt \lesssim 10$) in the localized phase, the memory should exhibit dynamical signatures that result from the distribution of effective couplings for the emergent local integrals of motion describing the localized bath.
For example, if the disorder distribution has sharp cutoffs, then this is manifest as oscillations in the memory.
These oscillations are damped out as $\gamma$ is increased, tuning the system and bath into the thermalized phase.
This behavior strongly depends on the distribution of disorder, as well as on the presence of bath-bath interactions.
Finally, at long times ($Jt \gtrsim 10$) we observe pathological exponential divergence of the memory kernel for certain realizations of disorder, deep in the thermalizing phase.
We find that this comes from exceptional points in the projected Liouvillian generating the dynamics of the memory kernel, which come about at real values of the coupling $\gamma$ due to the non-Hermiticity of the projection superoperator used to define the projected dynamics in the Nakajima-Zwanzig formalism.
Unlike past work~\cite{Wilkie2001a} that treated such exponential divergences as unphysical and should therefore be discarded, we have taken the view here that the divergences have a meaningful impact on the population dynamics.
We argued that after disorder averaging the memory kernel, such pathological behaviors should preclude any exponential decay of the memory.
Instead, we find that the tail of the memory is consistent with a power-law decay $\sim t^{-2-\zeta}$, where $0 < \zeta < 1$.
We find that this form still holds true for different distributions of disorder.
However, in the noninteracting bath case of $J=0$, the strictly power law decay appears to be replaced with an oscillatory component with a decaying amplitude that we find to be consistent with a power-law.
In the interacting ($J=1$) case, such a power-law ansatz allows us to extract estimates of the disorder-averaged infinite-time population of the central qubit, solely from finite-time simulations.
While such a procedure was shown in the past to work well when one could define a cutoff time for the memory kernel~\cite{Cohen2011}, here we have argued for the possibility that no cutoff time exists and demonstrated a proof-of-concept approach for extracting the infinite time populations in such a scenario.

In the model we have studied in this paper, we have taken the central coupling to scale to zero as $\gamma / L$, in accordance with Refs.~\cite{Ponte2017, Hetterich2018} which have argued for its necessity to perturbatively preserve localized eigenstates.
As a consequence, we have argued that there arises a separation of timescales between the population dynamics ($\tau_{p_0}$) and its associated memory kernel ($\tau_K$).
Should we repeat our arguments from \nnref{sec:short_times}{section~}{} with a central coupling scaling as $\gamma/L^q$, we find that these two timescales remain separated for $q > 1/2$, but coincide for $0 < q \leq 1/2$.
It is not clear whether such a separation of timescales--where $\tau_{p_0} \gg \tau_K$ as $L\to\infty$--is required for the preservation of localization.
Heuristically speaking however, having $\tau_{p_0} \gg \tau_K$ does not appear at first glance to be strong enough to preserve all aspects of MBL.
One of the dynamical hallmarks of MBL is a logarithmically slow spreading of entanglement, i.e.\ spins on sites $i$ and $i + L/2$ become entangled after a timescale $\sim \exp(L/2\xi)$ with $\xi$ being the localization length~\cite{Abanin2019}.
Based on our view of the system dynamics from the memory kernel, the interaction between these two sites mediated by the central qubit should proceed on a timescales growing as a power of $L$, which is much shorter than the dephasing time $\sim \exp(L/2\xi)$ and thus may accelerate the dephasing process responsible for the slow dynamics in the MBL phase.
However, it was noted in Ref.~\cite{Hetterich2018} that the central qubit at best facilitates a subextensive transport of magnetization which augments, but does not destroy, the logarithmic growth of bipartite entanglement.

Our work also raises tantalizing questions about possible connections between poles of the Laplace-transformed memory kernel and thermalization/delocalization.
To this end, some work~\cite{Heiss2012,Heiss1990,Stransky2018,Cejnar2007} has been done to connect the proliferation of exceptional points in non-Hermitian systems to the appearance of quantum phase transitions and chaos.
By focusing on a subpart of a closed system, we are forced to consider non-Hermitian \textit{Liouvillians} giving rise to exceptional points in the space of operators.
Explorations in this direction may benefit from insights from the physics of Feshbach resonances.
Of course, we are severely limited by the system sizes amenable to numerical studies, thus we are able to do little more than remark on the coincidences we observe. 

On the more practical side, we have demonstrated that there may be enough information from finite time dynamics to yield knowledge about long time limits, should they exist.
While we have only demonstrated the extrapolation to $t=\infty$ of the population of the central qubit, we should in principle be able to use the same memory kernel and the Nakajima-Zwanzig equation in \nnref{eq:scalar_memory}{(}{)} to extend the computed dynamics to longer times.
That this is even possible should not be too surprising, given that \nnref{eq:scalar_memory}{(}{)} when discretized over time gives the same form as the ansatz underlying linear prediction~\cite{Barthel2009,Schollwock2011}, a method widely used for extending dynamical calculations.
What we have shown in this work is that there may be more physical content in such a procedure than was previously appreciated.
To explore these ideas more thoroughly warrants careful attention, particularly in regard to stability and applicability, which we shall leave for future work.

Finally, we note that any possibility of a pathological memory kernel at real $\gamma$ can be erased by choosing to work with self-adjoint projection superoperators $\supop{P}$.
One may be interested in doing so, for example, in order to approximate system dynamics from low order, analytical expansions of the memory kernel. 
In that case it would be beneficial to know that the error introduced by the approximation is not exponentially divergent with time.
It is as yet unclear whether self-adjoint projectors necessarily yield improvements, since pathological behaviors can still occur for complex couplings $\gamma$ to limit convergence of na\"ive series expansions.
We note, however, that previous work~\cite{Fischer2007,Barnes2011} saw benefits from applying symmetry-adapted ``correlated projectors''--which, we should point out, are manifestly self-adjoint in Liouville space--to low order expansions of the memory kernel.
We leave clarification of this point for future work.

\section{Acknowledgments}
We are grateful to Amikam Levy, Sebastian Wenderoth, and Michael Thoss for useful discussions.
This research used resources of the National Energy Research Scientific Computing Center, a U.S. Department of Energy Office of Science User Facility operated under Contract No. DE-AC02-05CH11231.

\section{References}
\bibliographystyle{iopart-num}
\bibliography{references}




\end{document}